\documentstyle[12pt]{article}

\makeatletter

\advance \voffset by -1.0in
\advance \hoffset by -0.6in
\@addtoreset{equation}{section}
\def\section{\@startsection {section}{1}{\z@}{-3.5ex plus -1ex minus
 -.2ex}{2.3ex plus .2ex}{\large\bf}}

\def\be{\begin{equation}}
\def\ee{\end{equation} }

\def\bmphi{\mbox{$$\boldmath$ \phi $\unboldmath$$}}

\def\zsl{{\bf{\sf Z}} \hspace{-5pt}{\sf Z}\hspace{1pt}}

\begin{document}
\parskip 4pt
\begin{flushright}

DTP 95-29

hep-th/9506099

1 June 1995
\end{flushright}

\vspace{0.5cm}

\begin{center}
{\LARGE Skyrme-Maxwell Solitons  in (2+1) Dimensions}

\vspace{1cm}
\baselineskip 21 pt
{\Large
J. Gladikowski\footnote{e-mail: {\tt jens.gladikowski@durham.ac.uk}},
 B.M.A.G. Piette\footnote{e-mail: {\tt b.m.a.g.piette@durham.ac.uk}} and
 B.J. Schroers\footnote{e-mail: {\tt b.j.schroers@durham.ac.uk}}
\\
Department of Mathematical Sciences, South Road\\
Durham DH1 3LE, United Kingdom \bigskip \\}


{\bf Abstract}

\end{center}

\baselineskip 15 pt

{\small
\noindent
A gauged  (2+1)-dimensional version of the Skyrme model
is investigated. The gauge group is $U(1)$ and the dynamics
of the  associated gauge potential is  governed by a Maxwell term.
In this model  there are topologically stable soliton solutions
carrying  magnetic flux which is not topologically quantized.
The properties of  rotationally symmetric solitons of degree one
and two are discussed in detail. It is shown that the electric field
for such solutions is necessarily zero. The solitons' shape, mass
and  magnetic flux depend on the $U(1)$  coupling constant,
and this dependence is studied numerically from very weak
to very strong coupling. \bigskip\\

\noindent PACS number(s): 11.10.Kk, 11.10.Lm,  11.27.+d, 12.39.Dc }

\baselineskip 18 pt


\section{Introduction}

The  Skyrme model
is a generalized non-linear sigma model
in (3+1) dimensions \cite{skyrme1}. It has soliton solutions
which, after suitable quantization,  are models for
physical nucleons \cite{anw}.
The theory is invariant under the group
$SO(3)_{\mbox{\tiny iso}}$
of iso-rotations, and electromagnetism is introduced into the
model by  gauging a $U(1)$ subgroup of  $SO(3)_{\mbox{\tiny iso}}$,
 see \cite{Witten} for details.
The resulting fully coupled Skyrme-Maxwell system is
mathematically hard to analyze, but of considerable physical
interest: it is  here that one should compute
the Skyrme model's prediction for the proton-neutron
mass difference for example. In fact such a computation
(which necessarily involves quantum theory) was first  attempted
in \cite{dur} where the authors made various approximations
based on the smallness of the fine structure constant.
The result was a mass difference of $m_p - m_n =1.08$ MeV, which has
 the wrong sign (in nature the proton is $1.29$ MeV lighter than the
neutron).

In this paper we investigate classical properties of
a   (2+1)-dimensional
version of the  gauged Skyrme model.
The model, to be introduced  in section
\ref{model}, is a gauged version of the baby Skyrme model studied
in \cite{multi} and \cite{dyn}
and contains a dynamical abelian gauge field.
It has soliton solutions which are stable for topological
reasons and which carry magnetic flux. However,
the gauge symmetry is unbroken and
the solitons differ from the much studied flux tubes or vortices
in the abelian Higgs model in that
their magnetic  flux is not quantized.
Contrary to the situation in (3+1)-dimensional Skyrme-Maxwell
theory
it is possible to compute
certain  soliton solutions in our model explicitly
with moderate numerical effort,
and to investigate their structure quantitatively.
Thus we study the dependence of the magnetic
flux and the solitons' mass on the electromagnetic
coupling constant, assess the back-reaction of the
electromagnetic field on the matter fields and
make some  semi-quantitative statements about
the long range  inter-soliton forces.


\section{Gauging the Baby Skyrme Model} \label{model}


The model we want to study
  is defined on (2+1)-dimensional Minkowski space,  whose
signature we take  to be $(-,+,+)$. Points in Minkowski space
are written as $(t,{\bf x})$ or simply $x$, with coordinates
$x^{\alpha}$,   $\alpha = 0,1,2$, and the velocity of light is set to 1.
 We will mostly be concerned with
static fields and sometimes use the label $i=1,2$ for the coordinates of
the spatial vector {\bf x}.
The basic fields  are a scalar field  $\bmphi $ describing
matter and  a $U(1)$ gauge potential
$A_{\alpha}$ for the electromagnetic field. More precisely,
$\bmphi(x)$ is a  3-component vector
satisfying the constraint
$\bmphi\cdot \bmphi = \phi_1^2+\phi_2^2+\phi_3^2 =1$,
thus lying on a  2-sphere which we denote  by
${\cal S}_\phi^2$.

The model is a gauged version  of the baby Skyrme
model studied in detail in \cite{multi} and \cite{dyn}.
The Lagrangian considered there is invariant under global iso-rotations
of the field $\bmphi$ about a fixed axis ${\bf n} \in {\cal S}_\phi^2$.
Taking ${\bf n} = (0,0,1)$ for definiteness such a rotation can be
written in terms of the rotation angle  $\chi \in [0, 2\pi)$ as
\be
\label{isorot}
\left( \phi_1 ,\phi_2, \phi_3 \right)
\rightarrow (\cos \chi \: \phi_1 + \sin \chi \: \phi_2,
- \sin \chi \:\phi_1 + \cos\chi \:\phi_2, \phi_3)\,.
\ee
We   write $SO(2)_{\mbox{\tiny iso}}$ for the group of  such rotations.

Here we couple electromagnetism to the  baby  Skyrme model
by gauging the $SO(2)_{\mbox{\tiny iso}}$ symmetry. Thus
we require invariance under   local  rotations
\be
\bmphi \rightarrow O( x) \bmphi\,,\label{gauge}
\ee
where $O (x)$ is
an  $SO(2)_{\mbox{\tiny iso}} $ rotation matrix  which depends on $x$.
For infinitesimal  rotation angles $\varepsilon( x)$, this becomes
\be
\bmphi \rightarrow  \bmphi + \varepsilon {\bf n} \times \bmphi.
\ee
The abelian gauge field $A_\alpha$  transforms as
\mbox{$A_\alpha \rightarrow A_\alpha  - \partial_\alpha \varepsilon $},
so defining  the  covariant derivative  via
\be
 D_\alpha \bmphi = \partial_\alpha \bmphi
+ A_\alpha {\bf n} \times \bmphi,
\ee
one  has  \mbox{ $D_\alpha (O(x) \bmphi  ) =
O(x)D_\alpha \bmphi $}, as desired.
Finally  we define the curvature or field strength
$F_{\alpha \beta} = \partial_\alpha A_\beta - \partial_\beta A_\alpha$
with electric components $E_i = F_{i0}$ and the magnetic component
$B= F_{12}$. Thus
we can write down the Lagrangian  of our model:
\be
 L =
-H \int d^2x \left(
\frac{1}{2} ( D_\alpha \bmphi)^2 +
\frac{\lambda^2}{4} \left( D_\alpha \bmphi \times D_\beta \bmphi \right)^2 +
\mu^2 \left(1 - {\bf n \cdot \bmphi} \right)+
 \frac{1}{4g^2} F_{\alpha\beta}^2 \right)\,.\label{egg}
\ee
The first term is a gauged version of the $O(3)$ sigma
model Lagrangian (see \cite{raj}), the second is a gauged  Skyrme term,
the third  term may physically be thought of as a
pion mass term \cite{dyn} and the last term is the usual
Maxwell Lagrangian.
There are 4 free  parameters in this model.
$H$ has the dimension  energy, $\lambda$ and 1/$\mu$ are of
dimension length while $g$ represents the  coupling strength to the
gauge field and is also of dimension length.
We will discuss our choice of parameters in further detail below, but for
the time being we fix our energy  scale by setting $H= 1$.

It is  worth recalling  that the Skyrme term is  necessary in  the (ungauged)
Skyrme model  to prevent soliton solutions from collapsing to  singular
spikes. In the gauged model, however, the Maxwell term has the same scaling
behaviour as the Skyrme term, which suggests that there could be stable
solitons in a ``Skyrme-Maxwell theory without a Skyrme term".
This possibility is studied in \cite{Bernd}. Here we retain the Skyrme term
because we  are also interested in  properties of soliton solutions in
the limit of vanishing  electromagnetic coupling. The Skyrme term ensures
the existence of stable solitons in this limit.

The Euler-Lagrange equations for this model  can be written
conveniently in terms of
\be
\label{J}
{\bf J}_{\alpha} =
\bmphi \times D_{\alpha} \bmphi +\lambda^2  D_{\beta} \bmphi (D^{\beta}
\bmphi\cdot \bmphi \times D_{\alpha} \bmphi )
\ee
and  the conserved current
\be
\label{j}
j_{\alpha} = {\bf n} \cdot {\bf J}_{\alpha} \, .
\ee
They read:
\begin{eqnarray}
   D_{\alpha}{\bf J}^{\alpha} &=& \mu^2{\bf n}\times \bmphi \label{EL1} \\
   \partial_{\alpha} F^{\alpha \beta}&=& g^2 j^{\beta}\,. \label{EL2}
\end{eqnarray}
The second equation has three components which we want
to write  explicitly in terms
of the electric and magnetic field, for later use.
The $\beta = 0$ component is Gauss's law:
\be
\label{gauss}
\partial_i E_i = j_0.
\ee
The remaining two equations are particularly simple when expressed
in terms of polar coordinates
$(r,\theta)$ for $\bf x$.
Defining polar and radial coordinates of the current $j$ via
$j_{\theta} = x_1 j_2 -x_2 j_1$, $j_r =(x_1 j_1 + x_2 j_2)/r$
and analogously for the electric field $E_i$, we obtain
\begin{eqnarray}
\label{maxwell}
{\partial E_r \over \partial t} - \frac{1}{r} {\partial B \over
\partial \theta } &=& g^2 j_r \nonumber \\
{\partial E_{\theta} \over \partial t} + r{\partial B \over
  \partial r} &=& g^2 j_{\theta}\,.
\end{eqnarray}

\vbox{
The energy $E$ of a configuration $(\bmphi, A_{\alpha})$ is the sum
of the kinetic energy
\be
T = \frac{1}{2} \int d^2x \left( (D_0\bmphi)^2 + \frac {\lambda^2}{2}
(D_0 \bmphi \times D_i\bmphi)^2 +\frac{1}{g^2} E_i^2 \right)
\ee
and the potential energy
\be
\label{poten}
V = \frac{1}{2}\, \int d^2x \left( (D_1\bmphi)^2 +  (D_2\bmphi)^2
+\lambda^2 (D_1\bmphi \times D_2\bmphi)^2
+2\mu^2 (1-{\bf n}\cdot \bmphi) + \frac{1}{g^2} B^2\right).
\ee}
In this paper we are  only interested in  finite
energy  configurations,  so  we require that for all $t$
\be
\lim_{r \to \infty} \bmphi (t,{\bf x})
 = {\bf n} \label{pbound} \,.
\ee
This  boundary condition  allows the Euclidean space {\bf R$^2$} to be
compactified to a topological 2-sphere
${\cal S}^2_{\mbox{\scriptsize x}} $
 so that, at a given time $t$,  fields $\bmphi$ may
topologically be regarded as  maps
\be
\bmphi \quad :
\quad {\cal S}^2_{\mbox{\scriptsize x}} \rightarrow
{\cal S}^2_{\phi}\,. \label{pmap}
\ee
Such maps  are topologically classified by their degree $Q$,
which is an integer and can be calculated  via
\be
Q=\frac{1}{4\pi}
\int_{ {\cal S}^2_{\mbox{\scriptsize x }}} d^2 x \,
\bmphi\cdot\left(\partial_1 \bmphi
\times\partial_2\bmphi\right) \,.
\label{Q}
\ee
The degree is a homotopy invariant of  $\bmphi$ and therefore
conserved during time evolution.


\section{Static Solutions}


It is a well-known and important feature of the $O(3)$ sigma model
and its generalization to baby Skyrme models  that the potential
energy of a configuration is bounded below by
the modulus of its degree  (or a suitable multiple thereof).
 A similar result holds in our model, but
its proof requires a little work. In \cite{Bernd}, where the
the  potential  energy functional
\be
\label{aux}
V_{\mbox{\tiny aux}}[\bmphi,A_i] =  \frac{1}{2}\int
d^2 x\,   \left (( D_1 \bmphi)^2 + (D_2\bmphi)^2 +
( 1-{\bf n} \cdot \bmphi)^2 +
 B^2\right)
\ee
is studied in detail, it  is shown that $V_{\mbox{\tiny aux}}[\bmphi,A_i]$
is bounded below by $4\pi |Q|$.
Thus,  changing variables in  the expression for
$V$ via ${\bf x} \rightarrow \mu {\bf x}$, and discarding the
positive definite Skyrme term
we have the inequality
\be
V [\bmphi, A_i]\geq \frac{1}{2}\int d^2x \,  \left( (D_1\bmphi)^2
+ (D_2\bmphi)^2 +
2(1-{\bf n}\cdot \bmphi) + \frac{\mu ^2} {g^2 } B^2 \right).
\ee
Now note that, since $0 \leq (1-{\bf n}\cdot \bmphi) \leq 2$, it
 follows that
$(1-{\bf n}\cdot \bmphi) \geq \frac{1}{2}(1-{\bf n}\cdot \bmphi)^2$.
Thus we also deduce
\be
V [\bmphi, A_i] \geq \frac{1}{2} \int d^2x \, \left( (D_1\bmphi)^2
+  (D_2\bmphi)^2
+(1-{\bf n}\cdot \bmphi)^2 + \frac{\mu^2}{g^2 } B^2 \right).
\ee
If $\mu /g \geq 1$  it then follows immediately that
\be
\label{Bog1}
V [\bmphi, A_i]\geq
V_{\mbox{\tiny aux}}[\bmphi, A_i] \geq 4\pi |Q|.
\ee
If $\mu /g \leq 1$  on the other hand we have
\be
\label{Bog2}
V[\bmphi, A_i]\geq
 \frac{\mu^2} {g^2} V_{\mbox{\tiny aux}}[\bmphi, A_i] \geq 4\pi
 {\mu^2 \over  g^2} |Q|.
\ee
In both cases we have therefore  found a topological lower bound
for the potential  energy  $V$.

Our next goal is to find  static configurations of  given degree $n > 0$
which minimize the potential energy $V$.
To find these we  exploit the symmetries of
our model. Both  $V$ and $Q$ are invariant
 under
spatial rotations and translations
${\bf x} \rightarrow R{\bf x} + {\bf d}$, where
$R$ is an $SO(2)$-matrix and ${\bf d}$ a translation vector in {\bf R}$^2$,
and  under
global $SO(2)_{\mbox{\tiny iso}}$ rotations
defined earlier (\ref{isorot}).
They are also  invariant under simultaneous  reflections in the
Euclidean plane and on the ${\cal S}^2_\phi$ manifold:
\begin{eqnarray}
\label{reflect}
(x_1,x_2) &\rightarrow & (-x_1, x_2) \nonumber \\
\left( \phi_1 ,\phi_2, \phi_3
\right) & \rightarrow  & \left( -\phi_1, \phi_2, \phi_3 \right)\,.
\end{eqnarray}
Physically one may think of this  transformation as simultaneous
electric charge conjugation  and parity  operations.

Translationally invariant fields necessarily have degree zero,
but it is possible to  write down fields of arbitrary degree
which are invariant under the reflection
(\ref{reflect})
and  under
a combination of a rotation by some angle $\chi \in [0, 2\pi)$ and an
iso-rotation by $-n\chi$.
The appropriate ansatz for the scalar field $\bmphi$  is,
in terms of polar coordinates $(r,
\theta)$ for {\bf x},
\be
\label{hedgehog}
\bmphi ( r, \theta ) =
\left(
\begin{array}{rl}
\sin f(r) & \cos  n \theta   \\
\sin f(r) & \sin n \theta  \\
\cos f(r) &
\end{array}
\right) \,.
\ee
This is a two-dimensional version of the
 hedgehog ansatz used in the three dimensional Skyrme model.
Here
we consider configurations with their symmetry center, defined by
$\bmphi = -{\bf n}$,  at the origin.

 Under the reflection
(\ref{reflect}) the gauge field transforms as
\be
A_0 \mapsto -A_0, \qquad A_r \mapsto -A_r,
 \qquad A_{\theta} \mapsto A_{\theta},
\ee
 with the polar and radial coordinates of $A_i$ defined analogously to
those of $j_i$ before  Eq. (\ref{maxwell}).
Thus the requirement of rotational symmetry and reflection symmetry
implies the following form for the gauge potential
\be
A_0 = A_r =0\,, \qquad A_\theta = na(r)\, ,\label{ansatz}
\ee
where $a$ is an arbitrary function and the factor $n$ is
 introduced for convenience. For such a gauge field the electric field
vanishes, and the magnetic field is given by
\be
\label{magfield}
B = n{a' \over r}.
\ee

To ensure that the field is regular at the origin we  impose
\be
\label{origin}
f(0) = k\pi \, , \quad k \in \zsl  \quad \mbox{and} \quad
a(0) =  0 \, ,
\ee
and the finite energy requirement implies for the function $f$
\be
\label{inf}
\lim_{r \to \infty} f(r) = 0 .
\ee
With these boundary conditions the topological charge $Q$ of the hedgehog
configuration (\ref{hedgehog}) is equal to $-n$
if $k$ is odd and zero otherwise \cite{multi}. In  the following we will
restrict attention to $k=1$.

For configurations of the form (\ref{hedgehog}) and  (\ref{ansatz})
the current $j_{\alpha}$ has only one non-vanishing component,
namely $j_{\theta}$:
\be
\label{jtheta}
j_{\theta} = n (1+a)(1+ \lambda^2 f'^2)\sin^2 f.
\ee
The electric field vanishes and the magnetic field is independent
of $\theta$, so only the $\theta$-component of  the Euler-Lagrange
Eq. (\ref{EL2}) is non-trivial. Thus the field  equations (\ref{EL1})
and (\ref{EL2})
imply two  equations for $a$ and $f$, which read as follows:
\begin{eqnarray}
\label{eqf}
f^{''}\left( 1 + \lambda^2 \tilde{a}^2 \sin^2 f \right)
 +
\frac{f^{'}}{r} \left(\left( 2(\tilde{a}r)^{'} -
\tilde{a} \right) \lambda^2 \tilde{a} \sin^2 f + \lambda^2 rf^{'}\tilde{a}^2
\sin f \cos f + 1 \right) \nonumber \\
\qquad \qquad \qquad -\tilde{a}^2 \sin f \cos f - \mu^2  \sin f
 = 0 \, ,
\end{eqnarray}
where $\tilde{a} =  n(a+1)/r$,
and
\be
\label{eqa}
a^{''} -\frac{1}{r}{a^{'}}=g^2 (1+a)(1+\lambda^2 {f^{'}}^2)\sin^2 f \,.
\ee
Note  that $a=0$ is not a solution of  the second equation.
Since other constant solutions are forbidden
by the boundary condition (\ref{origin}) it follows
that all solutions will have a non-trivial magnetic
field given by (\ref{magfield}).

We will discuss the solutions of  (\ref{eqf})
 and (\ref{eqa}) in detail in the
next  sections, but first we want to address another question
of principal interest. Are there finite-energy solutions of the
field equations (\ref{EL1})  and (\ref{EL2})  which have a time-independent
purely radial  electric
field?
First we note
that, in two spatial dimensions,
 finite energy solutions necessarily have zero electric charge
\be
q= \int d^2 x \, j_0({\bf x}).
\ee
For  it follows from  Gauss's law (\ref{gauss})
that the modulus of the electric field falls off like $q/r$ for
large $r$. Hence the electric energy $\int d^2 x \, E^2_i$ diverges
if $q\neq 0$.
However, this argument does not rule out finite energy
solutions  with a non-trivial but spherically symmetric charge
distribution which integrates to zero. Like  for example
the hydrogen atom such a distributions would only produce a
short range electric field. We claim that this possibility
is not realized in our model.
  To prove this assertion we must allow  for more general fields
than considered so far.  In particular we can no longer impose the
reflection symmetry (\ref{reflect}) since it eliminates radial electric
fields from the start. Imposing  invariance under simultaneous spatial
rotations  and iso-rotations leads to the following general form
for the scalar field:
\be
\label{spiral}
\bmphi ( r, \theta,t ) =
\left(
\begin{array}{rl}
\sin f(r) & \cos  (n \theta -\chi(r,t))   \\
\sin f(r) & \sin (n \theta -\chi(r,t)) \\
\cos f(r) &
\end{array}
\right) \, ,
\ee
where $\chi(r,t)$ is an arbitrary function of $r$ and $t$.
However, having introduced this function we can  immediately
remove it by a gauge transformation which brings (\ref{spiral})
into the standard hedgehog form (\ref{hedgehog}). Thus having
fixed the gauge we write down the most general
gauge field which gives rise to a purely radial time-independent electric
field
\be
A_0 = v(r), \qquad  A_r = h(r)\,t, \qquad A_{\theta} = n a(r),
\ee
where $v$ and $h$  are arbitrary functions of $r$.
The electric field is  then
\be
E_i = -(v'(r) + h(r)){x_i\over r}.
\ee
Inserting this ansatz into the  field equation (\ref{EL1})  and
(\ref{EL2}) leads to   a complicated set of coupled differential  equations.
 Let us first consider the ``electromagnetic" equations (\ref{EL2}).
The $\theta$-component of the current $j_{\alpha}$ is still given by
(\ref{jtheta}), so the equation implied by the $\theta$-component
of (\ref{EL2}) is  (\ref{eqa}) as before. However,  both  the $t$ and
$r$ component of (\ref{EL2})  now  lead to non-trivial equations
which read
\begin{eqnarray}
t \,h\, \sin^2f &=& 0\label{h} \\
v^{''} + \frac{1}{r} v^{'} &=& g^2 v(1+{f^{'}}^2) \sin^2 f \label{v}  \,.
\end{eqnarray}
The first clearly implies that $h$ is identically zero.
To analyze the second we first note that $v$
has to satisfy the   boundary condition $v'(0)=0$ to  ensure that
the  electric field is regular at the origin and that for large
$r$,  $v'(r)$ has to  tend to 0 faster than $1/r$   for the electric
field energy to be finite. However, under these conditions
we can multiply (\ref{v}) by $(r\,v(r))$,
integrate both sides over $r$ from  0 to $\infty$ and
finally integrate
by parts  to obtain
\be
\int_0^{\infty} r  dr \,\left( (v')^2 + v^2\, g^2(1+\lambda^2 f'^2)\sin^2
f)\right)
=0\,.
\ee
Since the integrand is positive definite it follows that $v=0$
everywhere.
Thus $v=h=0$, and the functions $f$ and $a$ satisfy the same equations
as before. In particular, the electric field of the solution
vanishes everywhere.

%

\section{Asymptotic Properties}

%
To learn more about the minimal energy solutions of the
rotation  and reflection symmetric form
(\ref{hedgehog}) and (\ref{ansatz})  we need to
solve of the boundary value problem posed
by the coupled second order equations (\ref{eqf}) and (\ref{eqa})
and the boundary conditions
(\ref{origin}) and (\ref{inf}).
This requires  a careful analysis of the  equations near the
regular singular points
 $r=0$ and $r=\infty$  of
(\ref{eqf}) and (\ref{eqa}).

At the origin, $f$ and $a$ behave as follows
\begin{eqnarray}
f(r) &\approx& \pi + c_0 r^n + c_2 r^{n+2}\,\\
a(r) &\approx&  d_0r^2 +
g^2 \frac{c_0^2 \left( 1 + \delta^1_n  c_0^2 \right)}{4n(n+1)}
r^{2n+2}\, ,
\end{eqnarray}
where $\delta^1_n$ is the Kronecker symbol, $c_0$ and $d_0$
are arbitrary parameters
and $c_2$  is   a function of $c_0$, $d_0$, $n$  and $\mu$.

We already know that $f$  tends to zero  for large $r$
(\ref{inf}). Thus the Eq. (\ref{eqa}) becomes, for large $r$,
\be
ra^{''} = a^{'} \,.\label{aas}
\ee
which is solved by a constant function or by $a(r) = r^2$.
 Since  the  latter leads
to a magnetic field with infinite energy we conclude that there exist
a number  $a_{\infty}$ such  that
\be
\label{ainf}
\lim_{r\rightarrow \infty} a(r) = a_{\infty}.
\ee
Note that the finite energy requirement does not impose
any restrictions on  the value of $a_{\infty}$. Since $a_{\infty}$
is related to the
magnetic flux
\be
\label{flux}
\Phi = \int d^2 x \, B
\ee
via
\be
\label{fluxq}
\Phi = 2\pi n a_{\infty},
\ee
 there are also no {\it a priori}
restrictions on the value of the magnetic flux.
 This should be contrasted with
the situation in the ablian Higgs model, for example, where
the flux is quantized.

It follows from (\ref{ainf}) that the
 Eq.  (\ref{eqf}) can be simplified for
large $r$, and becomes
\be
f^{''}+\frac{f^{'}}{r}-
\left(\frac{n^2(a_\infty+1)^2}{r^2} + \mu^2 \right)f = 0 \,. \label{fas}
\ee
The solutions of this equation are the
modified Bessel functions  $K_m(\mu r), m = n(a_\infty +1)$.
Thus $f$ is  asymptotically proportional to
\be
K_m(\mu r) \sim
\sqrt{\frac{\pi}{2 \mu r}}e^{-\mu r}
\left(1 +{\cal O}\left( \frac{1}{\mu r} \right) \right) \label{fasymp}\,.
\ee
We then deduce from (\ref{eqa}) the asymptotic proportionality
\be
B(r) \propto -\frac{1}{r}e^{-2\mu r}\left(1 - {\cal O}\left(\frac{1}{2\mu
r}\right)\right)\,.\label{basymp}
\ee
This shows in particular  that the magnetic field has no long range
component.

Having understood the  asymptotic  properties of Eqs.
(\ref{eqf}) and (\ref{eqa}) it is relatively straightforward to
solve them numerically.
We have done  this for $n=1$ and $n=2$ using both a
 shooting method and a relaxation technique, with identical
results. The more general study of static multisoliton solutions in
the (ungauged) baby Skyrme model  in \cite{multi} suggests that minimal
energy configurations  have the rotationally
symmetric form considered here for degrees 1 and 2, but are less symmetric
for higher degrees. We expect that the rotationally symmetric solutions
of degree $n>2$  are similarly not true minima of the potential energy
in our gauged model, and we therefore  do not consider them here.

 The solution for $n=1$  corresponds to  the basic soliton
of our model, which we call a gauged  baby Skyrmion.
Already  the simple
asymptotic analysis carried out in this section tells us some
of its qualitative  features.
The profile function $f$ of the gauged baby Skyrmion has
the same asymptotic behaviour as that of the  the  ungauged baby
Skyrmion discussed in  \cite{multi}.
As explained there the resulting asymptotic forms of the fields $\phi_1$ and
$\phi_2$ are the same as those produced
by two orthogonal   {\it scalar} dipoles   in classical
Klein-Gordon theory. Thus, from the  point of view of the scalar (or matter)
fields, the gauged baby Skyrmion looks from afar like a pair of
orthogonal scalar dipoles in the plane of motion.

 In addition, however, the gauged baby Skyrmion
has a non-trivial electric current distribution with a magnetic
dipole  moment
orthogonal to the plane of motion.
Such a magnetic dipole moment
does not produce a  long range magnetic field in (2+1)-dimensional
electromagnetism, but it does carry magnetic flux.
Our soliton similarly has no long-range magnetic
field and also carries magnetic flux. Thus from afar
 a gauged baby Skyrmion looks like a triplet of
mutually  orthogonal dipoles: two scalar dipoles in the plane of motion and
one magnetic dipole orthogonal to it.

%
%
\section{Numerical Results}\label{res}
%
%

To compute explicit solutions
of (\ref{eqf}) and (\ref{eqa}) we need to fix the parameters of the model.
We can fix the energy and length scales  by setting  $H=1$ and  $\lambda =1$,
so that we are working in geometric units where all quantities are
dimensionless.
We further choose  $\mu^2 =0.1$
in order to  be able to compare our results with
 the discussion of the ungauged
baby Skyrme model in \cite{multi}
(where the reader  can read about the  motivation for this choice).
The electromagnetic coupling constant $g$, however, is not kept fixed.
In fact  we  are particularly interested in the  dependence of the
baby Skyrmion's properties on $g$. Consider first two global
properties: the  energy or mass, and the  magnetic
flux $\Phi$ (\ref{flux}).

 The  energy of the solutions for $n=1$ and $n=2$ is shown in
Fig. 1 as a  function of $g$.
When $n=2$, the  energy  is less than twice the energy of the $n=1$
soliton for all $g$, so this solution may be thought of as
 a bound state of two
gauged baby Skyrmions. Recall that the Bogomol'nyi bound on
the energy depends on whether $g \leq \mu $ or $g \geq \mu$.
In  the first regime the bound is  independent of $g$, but in
the second it decreases like $1/g^2$, compare (\ref{Bog1}) and  (\ref{Bog2}).
{}From   Fig. 1  it is clear that the energy  shows a very similar
dependence on $g$.
For  both $n=1$ and $n=2$  it is essentially constant in the first regime,
staying  approximately 50 \% above the Bogomol'nyi bound.
Here we find  in particular that in the limit $g\rightarrow 0$
the  energy of  the $n=1$ and $n=2$ solution tends to
$E_1 =  1.564\cdot 4\pi$ and $E_2 = 2.936\cdot 4\pi$ respectively,
which  agrees with the calculation  for  the ungauged model in \cite{multi}.
In  the second regime, by contrast,
  the energy, like the Bogomol'nyi bound,
decreases rapidly as $g$ is increased further.
However, our numerical results suggest
that for both $n=1$ and $n=2$
 the energy  tends to a non-zero limit  for $g\rightarrow \infty$.

The precise dependence of the magnetic flux
$\Phi$ on the coupling constant $g$ is shown
in Fig. 2 for both $n=1$ and $n=2$.  In the limit $g\rightarrow 0$
the magnetic flux tends to zero,
which is what one expects physically and which
 one can understand analytically by noting that
 in the limit $g\rightarrow  0$   Eq.
(\ref{eqa})  becomes  Eq. (\ref{aas}). As we saw in our
earlier discussion of that equation, the only finite energy solution
is the constant solution. It then follows from the boundary condition
(\ref{origin}) that $a$ and hence also $\Phi$  vanish in this limit.
Furthermore, integrating Eq.
 (\ref{eqa}) once and using $\lim_{r \rightarrow \infty} B(r) = 0$ we find
\be
B(0) = g^2 \int_0^{\infty} dr \, \left({1 + a \over r}\right)
 (1+ \lambda^2 f'^2)
\sin^2 f.
\ee
Numerically we observe
 that the dependence of $a$ and $f$ on $g$  is small for
weak coupling, and that  the flux
 $\Phi$ is approximately proportional to $B(0)$,
so we expect  $\Phi$ to grow quadratically with $g$ for small $g$.
This is precisely what the  double logarithmic plot in Fig. 2 shows:
there
is  a  weak coupling  range ($ 0 \leq g  \leq  \mu$)  where  $\log \Phi$
is a linear function of  $\log g$ with gradient $2$. Thus in this region
 the  magnetic flux of  the solutions of degree $n=1$ and $n=2$  is
approximately
\be
\Phi \approx -C_n g^2\,,
\ee
where $C_1\approx  24.5$ and $C_2\approx  31.5$.

For large $g$  the flux of the degree $n$ solution tends to
$-2 \pi n$ for  $n=1,2$. These  are precisely the
allowed values of
the magnetic flux in  models such as the abelian
Higgs model where the flux is quantized for topological reasons.
Thus, although there is no such
reason for flux quantization in our model we observe an effective
quantization in the strong coupling limit.

To understand
the effective flux quantization and the limit of the energy as
$g\rightarrow \infty$  better, we
 look at the dependence of  the
gauged baby Skyrmion's shape on the coupling $g$.
The function $a$ characterizing the magnetic field
is plotted in  Fig. 3(a) for several values of $g$.
Note in particular that for strong coupling, $a$ tends to
a step function, taking the value $0$ at the origin but $-1$
everywhere else.
Thus  at strong  coupling
the magnetic field  is  increasingly localized at the origin.
This is certainly consistent with Eq. (\ref{eqa}) in the
limit of large $g$, although it is not obviously implied by it.
Note also that $a_{\infty}=-1$  implies, via (\ref{fluxq}),
our earlier observation
that  at strong coupling the magnetic flux is quantized in units of $2\pi$.

In Fig. 3(b) we  plot  the profile function $f$
for a range of couplings. We have not included
more plots at weak coupling ($g<\mu$) because the profile
function barely changes in this regime.  Thus, as assumed
in the calculation of the proton-neutron mass difference in \cite{dur},
the back reaction of  the electromagnetic field on the scalar field
is  negligible at weak coupling.
At strong coupling, however, the profile function changes
significantly, and  the  configuration becomes more localized.
We conjecture that $f$ also tends to a singular step function in the limit
$g\rightarrow \infty$. To justify this conjecture, consider
Eqs. (\ref{eqf}) and (\ref{eqa}). When $a$ is the step function
described above, these equations decouple everywhere except at
the origin, and the first equation becomes the Euler-Lagrange
equation derived from the  functional
\be
F[f] = \int dr \, r \left( \frac{1}{2} f'^2 + \mu^2(1-\cos f)\right).
\ee
However,  this functional cannot be minimized by a non-singular function
satisfying the boundary conditions (\ref{origin}) and (\ref{inf}).
For suppose there were such a minimum, and call it $f_1$. Then define
$f_{ \kappa }(r) = f_1 (r /  \kappa )$, and find that $F[f_{\kappa}] <
F[f_1]$ for $\kappa <1$. Thus one can always lower the value of $F[f]$
by making $f$  spikier.

To sum up, we have the following picture  for the $n=1$ and the
$n=2$ solitons in the strong coupling limit:  both the magnetic field
and the energy  distribution become localized near the origin, tending
to singular  distributions as $g\rightarrow \infty$.
In this limit, the total energy does not vanish because
 of the contributions from gradient terms in the energy density, and
the magnetic flux tends
to $-2\pi n$.

To end, let us  look at the solutions of (\ref{eqf}) and (\ref{eqa})
for a particular value of $g$ in more detail. Ideally we would
like to pick a ``physical" value for $g$ which
allows us to estimate the magnitude  of the effects we  have observed
in the  physically relevant (3+1)-dimensional Skyrme model.
In practice  it is not clear how to define ``physical" here,
but  the following procedure should
lead to a value
for $g$ which is at least in the  right  ball park.

The basic idea is to treat our model as if its solitons  described
physical baryons and its elementary quanta were physical pions:
this allows us to fix the  energy and length scale, and
 to compute a definite value for the  coupling constant $g$ from
 the physical value  of the fine structure constant.
Thus we  assume, in accordance with nature, that the electromagnetic
coupling constant is small.

First we fix the energy scale $H$ by identifying the mass of a  gauged
baby Skyrmion
with the physical  nucleon mass of $940$ MeV.
 Since for small $g$ the mass of the
gauged baby Skyrmion  is virtually independent of $g$, we  pick the
value of the mass at $g=0$; this leads to
$H = 48$ MeV.
To find a physical length scale $\lambda$ we  note that
$1/\mu$ is the equivalent to the Compton wavelength of the
pion in the Skyrme model.
 Thus we  choose $\lambda$ such that  $1/\mu = 1.41$ MeV,
i.e. $\lambda =\sqrt{0.1} \cdot 1.41 $fm $ =0.45 $ fm.
To compute Planck's constant in geometric units  we write
\be
\hbar = 197.3 \mbox{ MeV  fm} \approx 9.1 (48 \mbox{MeV})(0.45 \mbox{ fm})
\ee
and deduce that  $\hbar = 9.1$ in geometric  units.
Finally we use the physical value of the fine structure constant
$\alpha = e^2/(4\pi\hbar) \approx  1/137$. Here $e$  is the electron's
charge which is related to the coupling constant $g$ via $e=g\hbar$.
With $\hbar =9.1$ we   conclude $g=0.1$ in geometric units.

At this value for $g$  the gauged baby Skyrmion is
lighter than the baby Skyrmion at $g=0$
by $\Delta E_1 = 0.12$, which is $5.9$ MeV  in physical units.
For the solution with $n=2$  the corresponding energy
difference is
$\Delta E_2 = 0.56$, which is  $27$ MeV in physical units.
It is also interesting  to look at
the $g$-dependence of the  difference  between the  energy of the $n=2$
solution and twice the
energy of the $n=1$ solution, which  may be interpreted as a   binding energy.
In the $g=0$ case, this   is
about $6.6\%$, but it is $14\%$ when $g=0.1$.  Thus the inclusion
of the  electromagnetic
field  leads to a more strongly bound $n=2$ soliton.

The precise  shape of the solutions for $g=0.1$ can be seen
in Fig. 4., where we  plot the energy density  and the
magnetic field   for both $n=1$ and $n=2$.
The  baby Skyrmion's energy distribution
 is  bell-shaped and peaked at the  origin,
 whereas
the energy distribution for the soliton of degree 2 is maximal
on  a ring
with radius  $r = 1.78$.


\section{Conclusions}


In this paper we have studied soliton solutions of the
coupled Skyrme-Maxwell system in (2+1) dimensions.
The rotationally symmetric solitons  we have considered
necessarily carry a magnetic field but the  electric field is zero.
The magnetic flux can take arbitrary values, but in
the strong coupling limit we observe an effective flux
quantization.  The soliton  mass  decreases when the electromagnetic
coupling constant is  increased and all the other
parameters of the model are kept fixed.  Thus  a baby Skyrmion
can lower its mass by interacting with the electromagnetic field.

Although the $U(1)$ gauge group is unbroken the
baby Skyrmions' magnetic field  is short-ranged.
The reason for this is that the electromagnetic
current carried by the baby Skyrmion only has a
magnetic dipole component; in (2+1) dimensions
static magnetic dipoles, however,  have no
long-range fields  in Maxwellian electromagnetism.
This observation has important consequences for
the interaction of gauged baby Skyrmions.
Since the
scalar fields fall off like exp$(-\mu r)$ and the magnetic
field  like exp$(-2\mu r)$,  the magnetic forces will
be negligible compared to the scalar forces between
two well-separated gauged baby Skyrmions. Thus, to
first approximation, the forces should be the
scalar dipole-dipole  forces between purely scalar
baby Skyrmions discussed  in detail in \cite{dyn}.

\vspace{1cm}

\noindent {\large {\bf Acknowledgements}}

\noindent We are  grateful to  Wojtek Zakrzewski for stimulating discussions.
BJS acknowledges an SERC research assistantship
and JG  an  EPSRC studentship
no. 94002269.

\pagebreak
\parindent 0pt

{\large \bf Figure Captions}

\vspace{1.5cm}

{\bf Fig. 1}:
 Energy   ``per Skyrmion" as a function
of the coupling constant $g$ for $n=1$
and $n=2$ .  The dashed line  is a plot of the Bogomol'nyi bound, see
(\ref{Bog1}) and (\ref{Bog2}).

 \vspace{1cm}

{\bf Fig. 2}:
The magnetic flux  ``per Skyrmion" as a function of
 the coupling constant $g$ for $n=1$ and $n=2$.

\vspace{1cm}

{\bf Fig. 3}:
 The function $a$ ({\bf a}) and the profile function $f$ ({\bf b}) of the $n=1$
solution
for various values of the
coupling constant $g$.

\vspace{1cm}

{\bf Fig. 4}:
 The energy density ({\bf a}) and the magnetic field ({\bf b})
for $n=1$ and $n=2$ at $g=0.1$;
the function $e$  plotted in ({\bf a}) is
the integrand of (\ref{poten})  divided by $4\pi$.

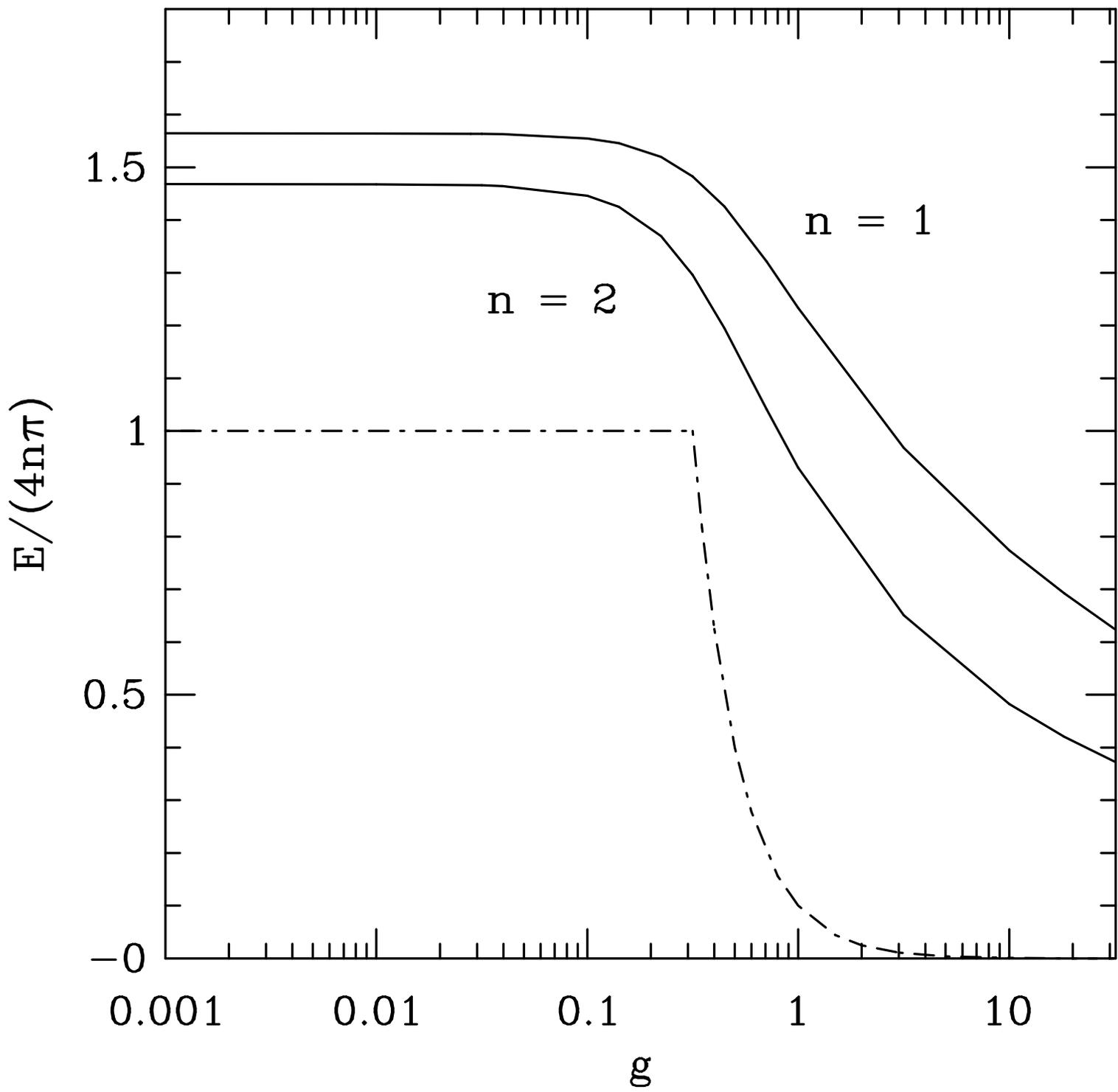
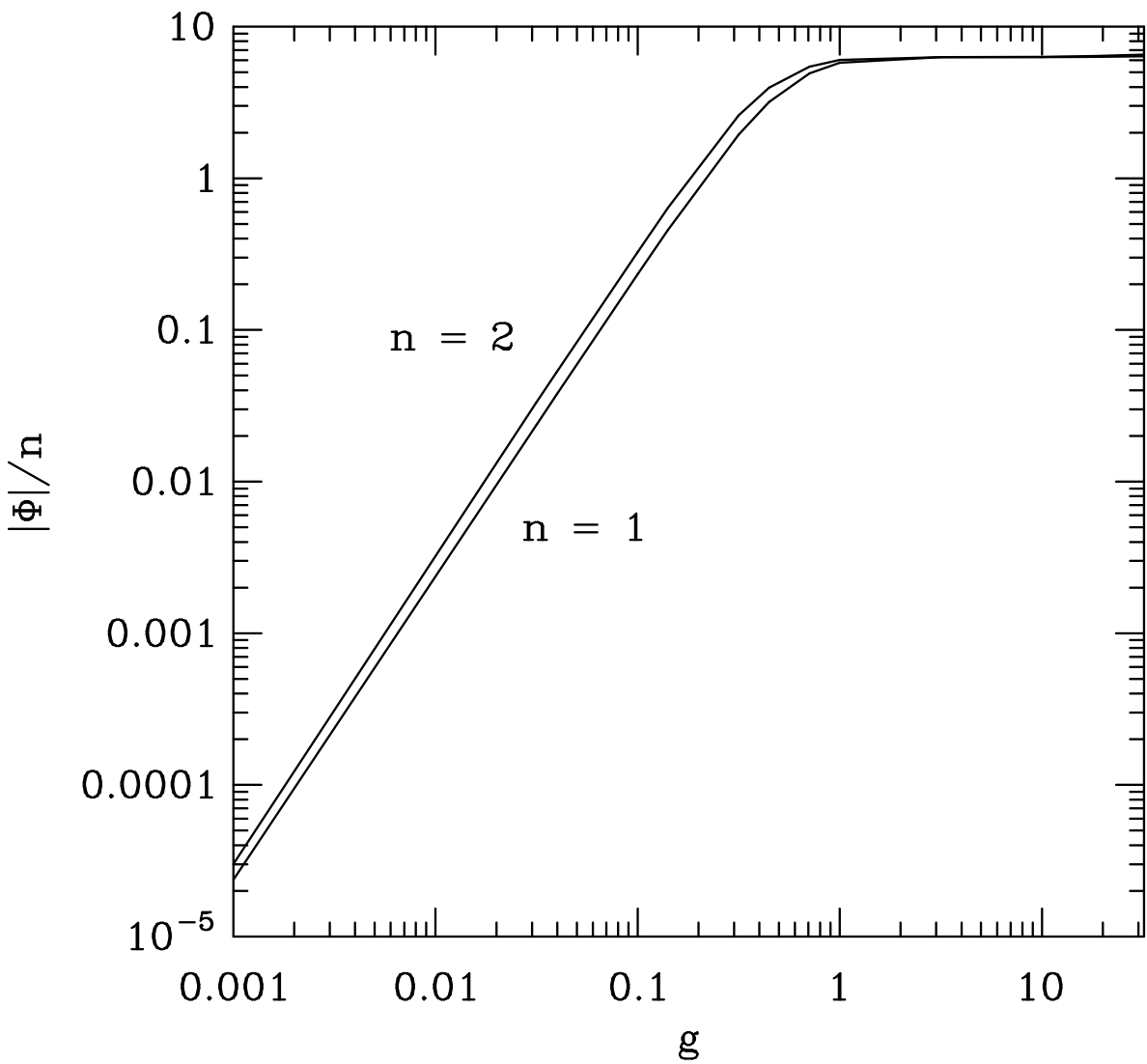
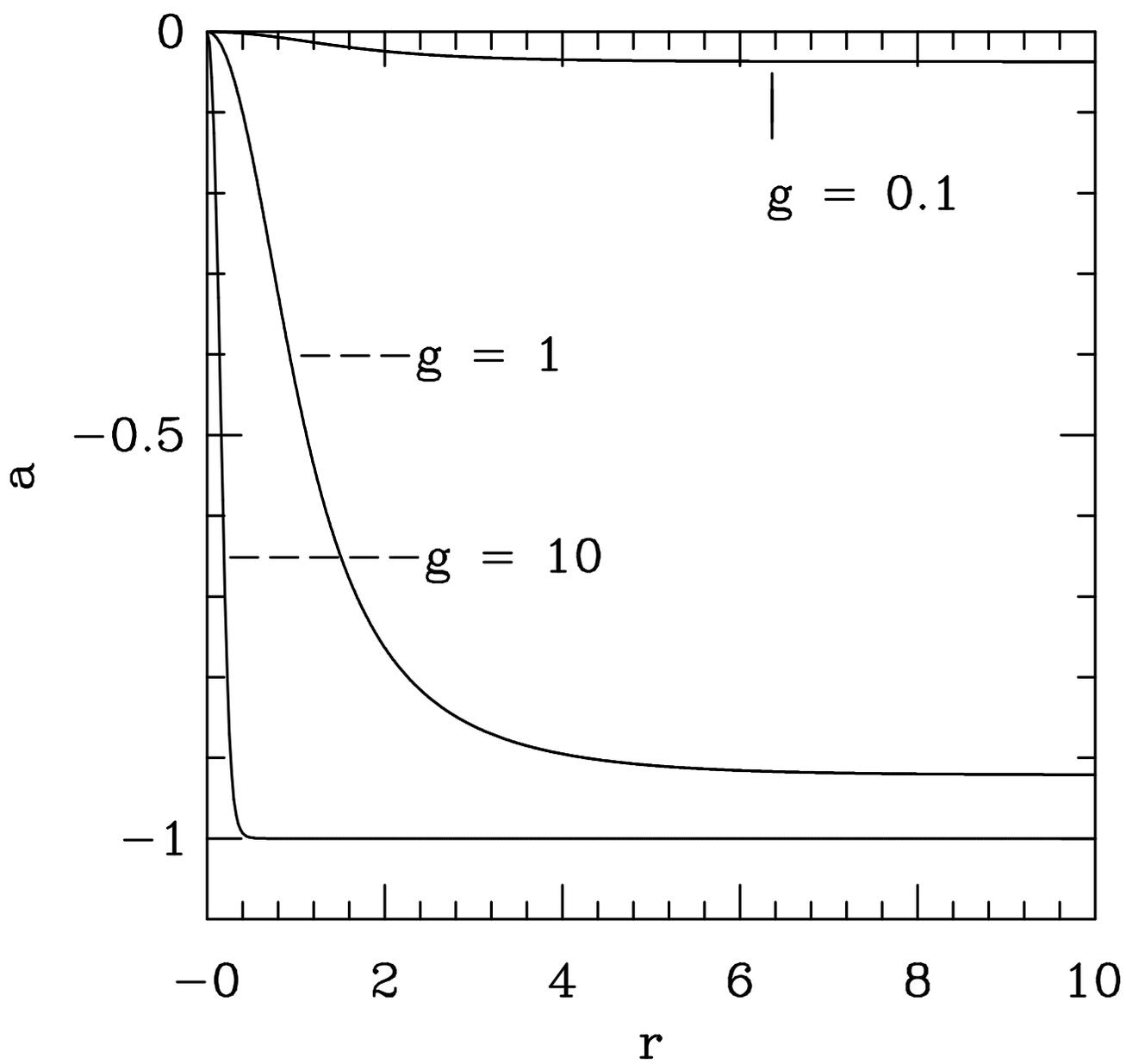
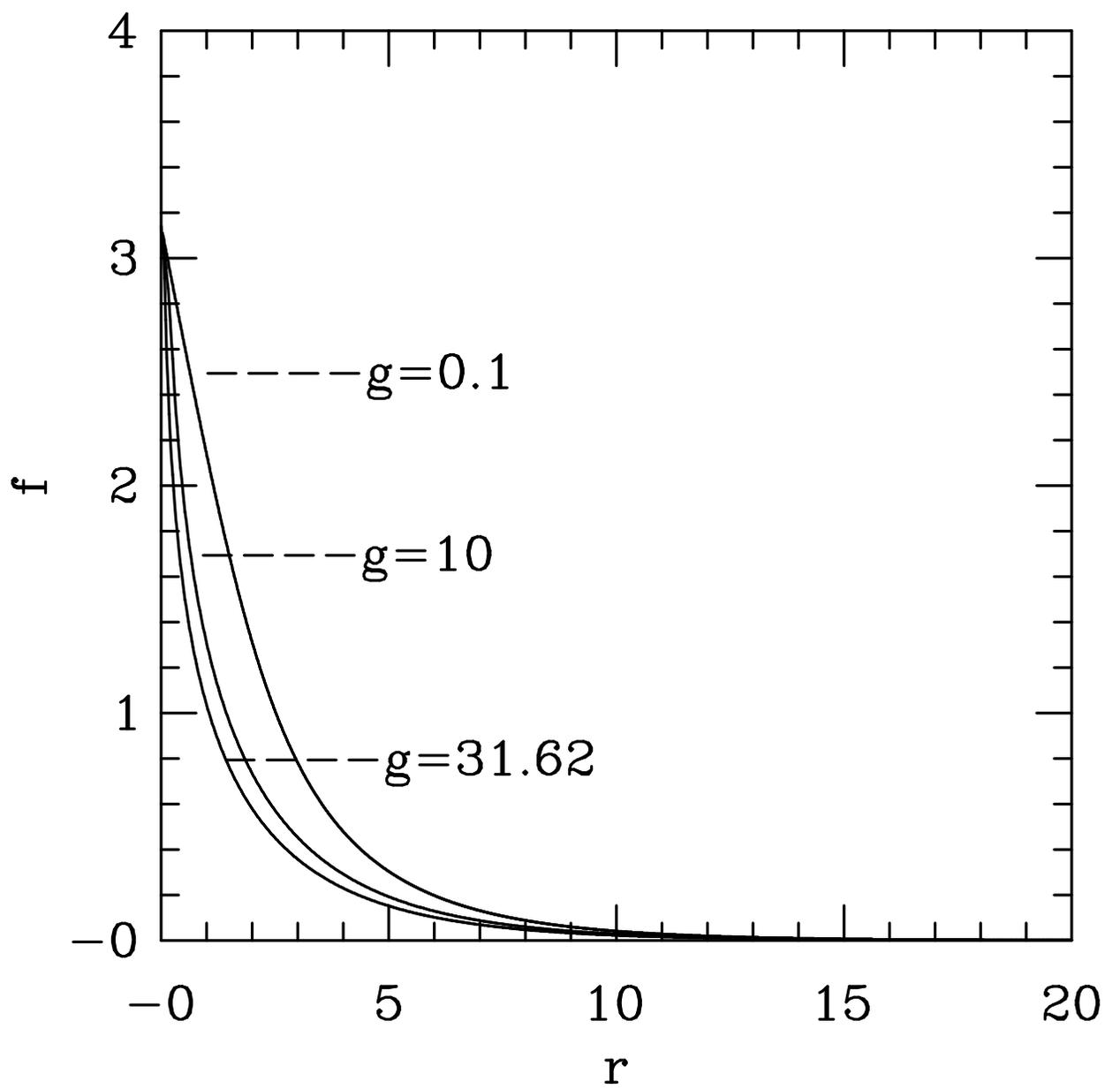
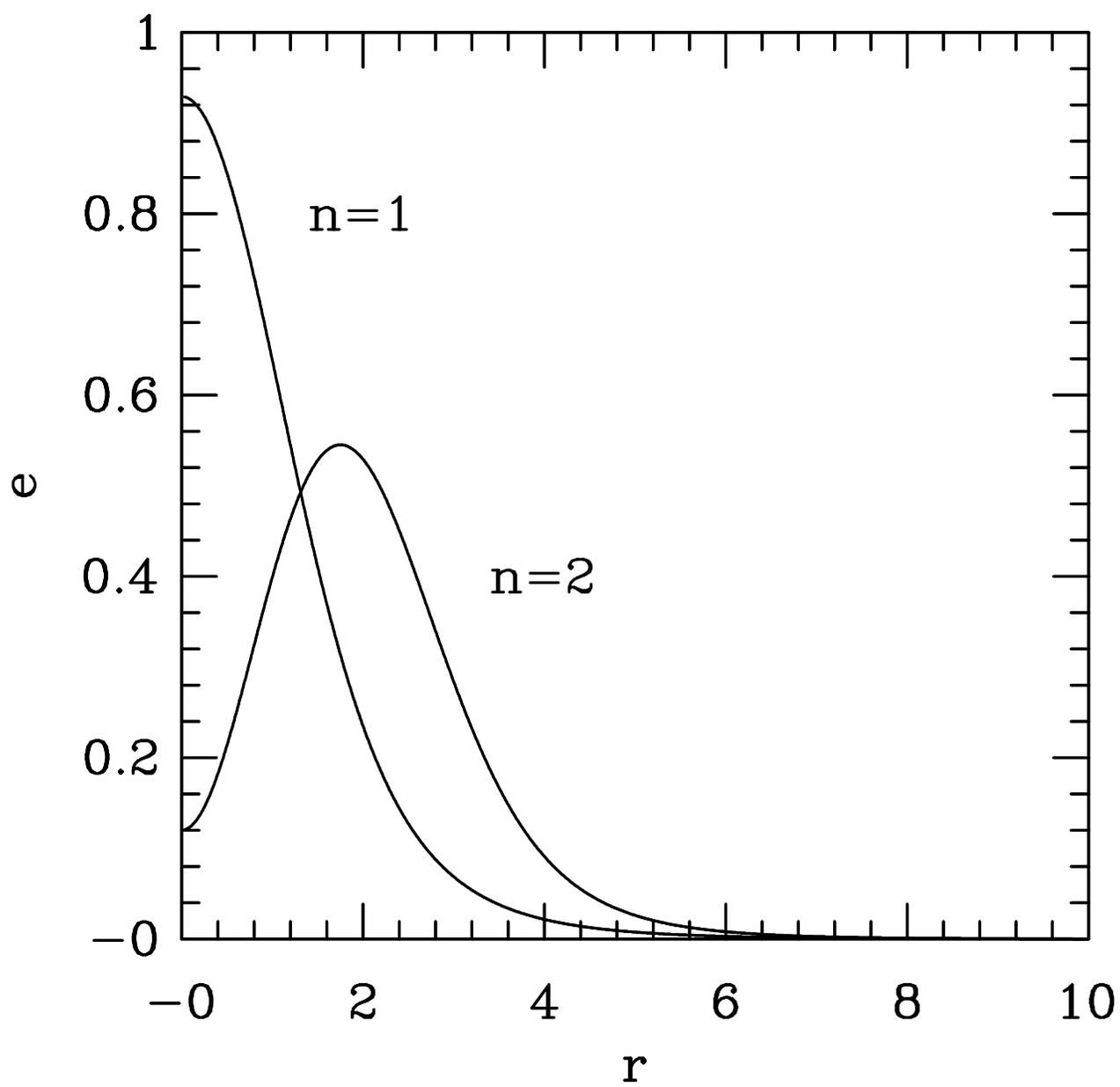
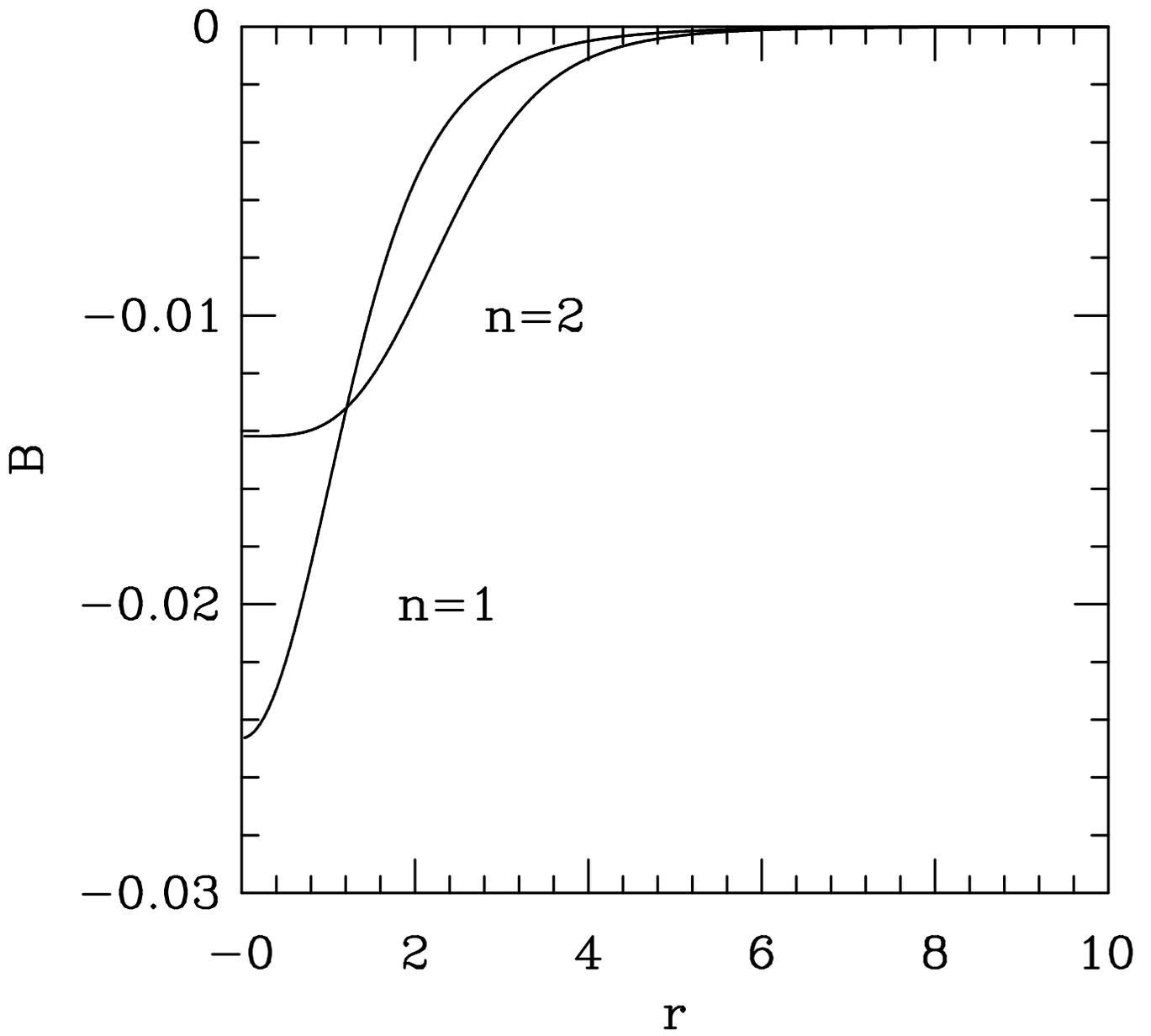
\end{document}